\RequirePackage{amsmath}
\documentclass[runningheads]{llncs}
\usepackage[T1]{fontenc}
\usepackage{graphicx}
\graphicspath{{figures/}}
\usepackage{color}
\usepackage{comment}
\usepackage{enumitem}
\usepackage{listings}

\usepackage{hyperref}
\hypersetup{colorlinks,allcolors=black}

\begin{document}

\title{A GPU-accelerated Molecular Docking Workflow with Kubernetes and Apache Airflow\thanks{Preprint submitted for publication.}}
\titlerunning{A Molecular Docking Workflow with Apache Airflow}
%
\author{Daniel Medeiros \and Gabin Schieffer \and Jacob Wahlgren \and Ivy Peng}
%
%
\institute{Department of Computer Science, KTH Royal Institute of Technology, Sweden\\\email{\{dadm, gabins, jacobwah, ivybopeng\}@kth.se}}

\maketitle              
\begin{abstract}
Complex workflows play a critical role in accelerating scientific discovery. In many scientific domains, efficient workflow management can lead to faster scientific output and broader user groups. Workflows that can leverage resources across the boundary between cloud and HPC are a strong driver for the convergence of HPC and cloud. This study investigates the transition and deployment of a GPU-accelerated molecular docking workflow that was designed for HPC systems onto a cloud-native environment with Kubernetes and Apache Airflow. The case study focuses on state-of-of-the-art molecular docking software for drug discovery. We provide a DAG-based implementation in Apache Airflow and technical details for GPU-accelerated deployment. We evaluated the workflow using the SWEETLEAD bioinformatics dataset and executed it in a Cloud environment with heterogeneous computing resources. Our workflow can effectively overlap different stages when mapped onto different computing resources.

\keywords{Cloud and HPC Convergence \and Workflow \and Kubernetes \and Apache Airflow \and Molecular Docking}
\end{abstract}

\section{Introduction}
The convergence of cloud and high-performance computing (HPC) is emerging to meet the increasing demands of diverse workloads. Major cloud providers, like Amazon Web Services (AWS), now provide on-demand availability to high-end computing capability on HPC hardware. On the other hand, many cloud techniques that have been matured by the community efforts for elastic executions, fault tolerance, virtualization, and isolation, are being explored by HPC users and systems to meet the increasing demands of HPC workloads. Complex workflows are a strong driver that can benefit from efficient management across HPC and cloud boundaries to ease the barrier to reaching wider user communities. For instance, cloud storage is often used for observation data for its high availability, and high computing power from HPC is used for compute-intensive high-fidelity simulations.

Workflows are built in many scientific domains to accelerate scientific discovery. Scientific workflows often are built to efficiently connect and coordinate tasks from data generation to pipelined data processing and analysis. For instance, massive data generated from Large Hadron Collider (LHC) experiments in CERN are consumed by numerous scientific analysis procedures built by domain scientists worldwide~\cite{hasham2011cms}. In medicine, molecular docking workflows are used for drug discovery~\cite{Venkatraman2022}. In structural biology, workflows of multi-scale simulations and machine learning methods have furthered the understanding of the structure of viruses~\cite{Trifan2022}. 
 
Workflows can be represented as directed-acyclic graphs (DAGs) to capture task dependencies. Depending on the support for different containers and schedulers, we can mainly classify workflow management software as either cloud-native if they support a container-based environment on a Kubernetes cluster or HPC-native if they are designed to be deployed on HPC clusters using Slurm-like schedulers. In this work, we explore using Apache Airflow workflow management software to support a molecular docking workflow that was designed for HPC systems in a container-based cloud-native environment. Our study is based on an important scientific application for drug discovery -- virtual screening using molecular docking software on GPU-accelerated compute nodes. For the case study, we use AutoDock-GPU, a state-of-the-art molecular docking software, where previous works have built a workflow for HPC cluster~\cite{legrand_gpu-accelerated_2020} and have been further accelerated with hardware features~\cite{schieffer2023tcu}. Our workflow decouples I/O intensive phases~\cite{legrand_gpu-accelerated_2020,markidis2021understanding} from compute-intensive phases and maps them to different hardware resources for improved resource utilization. To our best knowledge, this work is the first study exploring Apache Airflow in supporting such workflow on Cloud.

{}

In summary, we made the following contributions in this work:
\begin{itemize}[noitemsep,topsep=0pt]
    \item We analyze an HPC-targeted virtual screening workflow and identify opportunities for porting it to a Cloud environment;
    \item We design our workflow in terms of tasks, dependencies, and resources requirements and provide a DAG-based implementation in Apache Airflow;
    \item We evaluated our workflow implemented with Kubernetes and Apache Airflow and executed in a Cloud environment with heterogeneous computing resources.
\end{itemize}

\section{HPC and Cloud Workflows}
\label{sec:bg}
We compare the execution environments, tools and common practices between Cloud and HPC in this section. Table~\ref{tab:cloud_hpc} summarizes their main differences. We also describe the Apache Airflow workflow management software.
\begin{table}[bt]
    \centering
    \caption{Comparison of Cloud and HPC, with typical design choices.\label{tab:cloud_hpc}}
    \begin{tabular}{l|c|c}
        \hline& HPC & Cloud \\
        \hline\hline
        Allocation strategy & static & dynamic/on-demand \\
        Unit for allocation & node-granularity & fine-grained \\
        Execution & bare metal & containerized/virtualized \\
        Scheduler & SLURM & Kubernetes \\\hline
    \end{tabular}    
\end{table}

\subsection{Cloud and HPC Environments}
The paradigm for resources allocation greatly differs between cloud and HPC infrastructures. HPC systems usually rely on large computing clusters, where users request a particular amount of computing nodes, for a limited amount of time. The characteristics of each node are generally fixed, and cannot be tuned to the application specific requirements -- for example CPU type, or presence of hardware accelerators such as GPUs. In addition, infrastructure-related characteristics, such as storage and internode communication, are usually not modifiable by the user. In the cloud, resource allocation is finer, where sub-node allocation is common, and the user can request various types of resources based on their usage.\\
\textit{Opportunity 1: The fine-grained allocation in cloud allows tailoring resource allocation to specific workload needs, and thus improve resource utilization over the entire lifespan of the workload.}
\vspace{.5em}

HPC users usually execute their workloads on a bare metal environment, with highly-tuned execution environments, specifically tailored to the underlying hardware. This approach allows reaching high efficiency on a given system. However, this comes at the cost of reduced portability, as porting an application to another system may require some adaptation in the code and/or the system. A common solution to this lack of portability, widely used in cloud computing, is using \textit{containers}. A \textit{container} bundles one or more applications with a set of dependencies, and can be executed on various systems. This improves portability and usability of an application, and simplifies the co-location of several workloads on a single system. This technology is particularly used in cloud environments. \\
\textit{Opportunity 2: Containerizing workloads improves portability, and usability, but at the cost of formalizing application requirements.}
\vspace{.5em}

SLURM, PBS, Torque and Cobalt are popular schedulers in the HPC environment. In this context, any workflow management software targeting HPC environments must integrate with these schedulers in order to execute workflows seamlessly. In a cloud environment, the management of resources can be handled by an orchestrator, which configure and maps computing resources for use by a user workload. Kubernetes is a widely-used open-source orchestrator. It manages containerized workloads, by providing a standardized way of describing workload resources and requirements. 
Kubernetes provides auto-scaling capabilities, where the amount of computing resources can be automatically adapted as the application is running based on its needs.\\ 
\textit{Opportunity 3: Using cloud-native orchestration techniques to deploy our workload can improve portability, simplify deployment, and provide elasticity.}
\vspace{.5em}

{}



\subsection{Apache Airflow Workflow Management Software}
\begin{figure}[bt]
    \centering
    \includegraphics[width=0.6\textwidth]{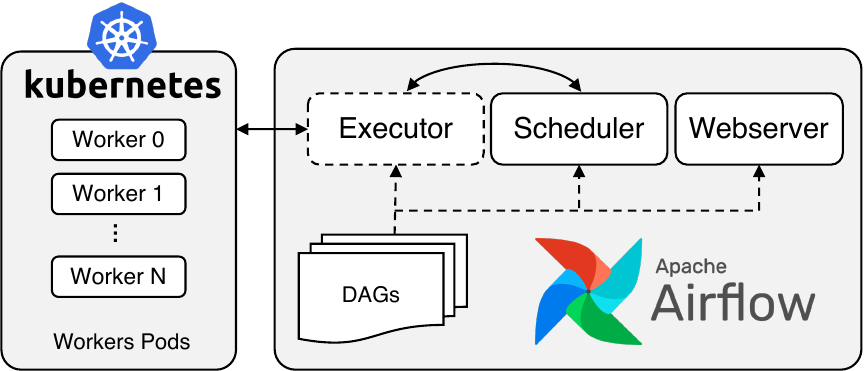}
    \caption{General architecture of Apache Airflow, and its interaction with a Kubernetes cluster. Arrows represent communication between components.}
    \label{fig:airflow}
\end{figure}

\noindent\textbf{DAG.} Workflows can be described as Directed Acyclic Graphs (DAG), where tasks are represented as nodes, and dependencies between tasks are represented as edges. This formal description of a workflow has several advantages. For instance, the DAG describing a workflow being a mathematical object, graph algorithms can be directly used to achieve specific goals without requiring algorithmic adaptation on a per-workflow basis. A relevant example in this work is efficient task scheduling. A DAG-aware task scheduler can be used instead of a customized user-built scheduler to achieve efficient scheduling of tasks, while enforcing requirements such as task dependencies.\\

\noindent\textbf{Apache Airflow}\footnote{https://airflow.apache.org/} is an open-source workflow management platform, where workflows are described as DAGs using the Python programming language. Airflow provides a simple way of efficiently scheduling and executing tasks from DAG-represented workflows on a Kubernetes cluster, which is leveraged in this work. Fig.~\ref{fig:airflow} describes the architecture of Airflow. The user creates a DAG as Python code, which is interpreted by Airflow and presented in a graphical web interface. The user can then interact with Airflow using this interface, for example, to trigger the workflow execution or to check workflow execution state. Once a workflow is triggered by the user, the tasks are automatically scheduled by Airflow, and executed on a Kubernetes cluster as soon as all task dependencies are fulfilled.

\noindent The following Airflow features are used in our work: \begin{itemize}[noitemsep,topsep=0pt]
    \item Defining a single DAG, along with a Docker image specific for our workflow, which provides the portability of a workflow; 
    \item Delegating task scheduling and execution to Apache Airflow, which results in a concise workflow description, and efficient task scheduling;
    \item Monitoring tools and visualization, which could be used by domain scientists.
\end{itemize}

\section{A virtual screening workflow on Apache Airflow}
In this section, we detail our design and implementation of a workflow for large-scale GPU-accelerated virtual screening in the Cloud. We first introduce AutoDock-GPU and then present an elementary molecular docking workflow. We then extend this elementary workflow into a large-scale virtual screening workflow on Apache Airflow.

AutoDock-GPU is a state-of-the-art GPU-accelerated molecular docking application. It is a variant of AutoDock~\cite{morris_automated_1998}, one widely used family of software for molecular docking simulations. Molecular docking methods are widely used in the pharmaceutical industry to characterize the ability of candidate drug molecules to bind themselves to identified targets in the human body, and therefore trigger their pharmacological effect. The main challenge in drug discovery is to be able to efficiently evaluate several millions of drug candidates, referred to as \textit{ligands}, against a single identified \textit{protein} target. This large-scale process is called \textit{virtual screening}.

All AutoDock variants use an energy-based scoring function to measure the quality of a given binding pose, i.e. the geometrical conformation of the ligand, this function is evaluated many times for each ligand-protein complex, and incurs high computational cost. Thus, offloading the compute-intensive part onto GPU has achieved orders of magnitude of speedup. 
A GPU-accelerated version of AutoDock has been developed under the name of AutoDock-GPU~\cite{santos-martins_accelerating_2021}. 
Its CUDA implementation~\cite{legrand_gpu-accelerated_2020} has been successfully used to perform large-scale screening of millions of drug candidates on the Summit supercomputer~\cite{legrand_gpu-accelerated_2020}. In this work, we use the CUDA version of AutoDock-GPU to build a virtual screening workflow. A previous workflow targeting HPC systems using Slurm has been developed~\cite{ornl_workflow}. It is worth noting here that since AutoDock variants share similar characteristics, the workflow we describe could be generalized to other AutoDock variants than AutoDock-GPU.


\subsection{An elementary molecular docking workflow}
\begin{figure}[bt]
        \centering
        \includegraphics[width=0.4\textwidth]{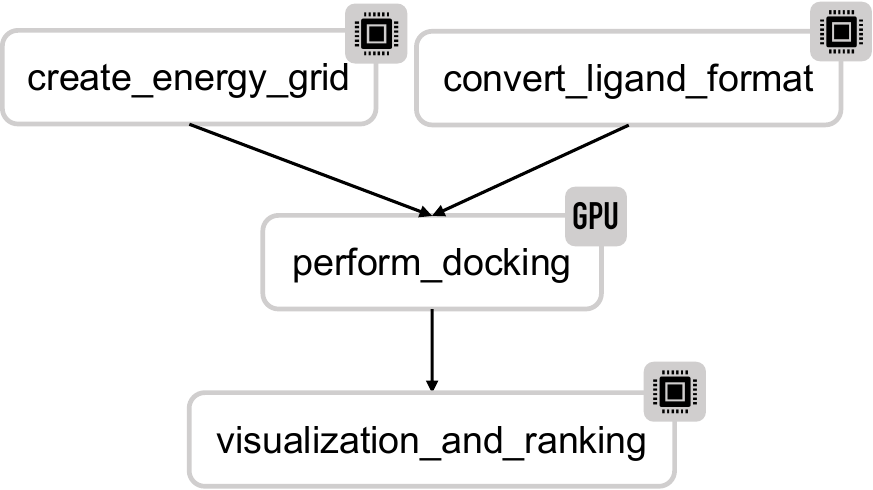}
        \caption{The DAG of an elementary molecular docking workflow: a single ligand is docked onto a single receptor. Resource requirements for each task is either GPU or CPU.}
        \label{fig:workflow_single_docking}
\end{figure}

To design our virtual screening workflow, we first studied the data requirements and task dependencies involved in a molecular docking job, between a single ligand and a single receptor. This \textit{elementary} workflow is presented in Fig.~\ref{fig:workflow_single_docking} as a DAG. The first takeaway from this workflow is that we can split the set of tasks into two categories. The first category is I/O-related tasks, where file reading, conversion, and writing are performed. This category of tasks only require limited CPU resources to execute, tasks within this category are depicted with a CPU icon in the DAG. The second category of tasks, which represents the main computational cost of this workflow, is the docking tasks, which require GPU resources atop CPU, to accelerate the process.


\subsection{A large-scale virtual screening workflow}
\begin{figure}[bt]
    \centering
    \includegraphics[width=.6\textwidth]{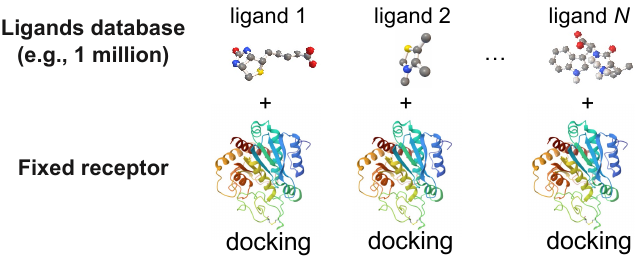}
    \caption{A virtual screening process, where a single protein receptor is identified beforehand (Fixed receptor), and millions of ligand molecules are evaluated against the receptor using molecular docking methods.}
    \label{fig:docking_is_expensive}
\end{figure}
\begin{figure}[bt]
    \centering
    \includegraphics[width=.6\textwidth]{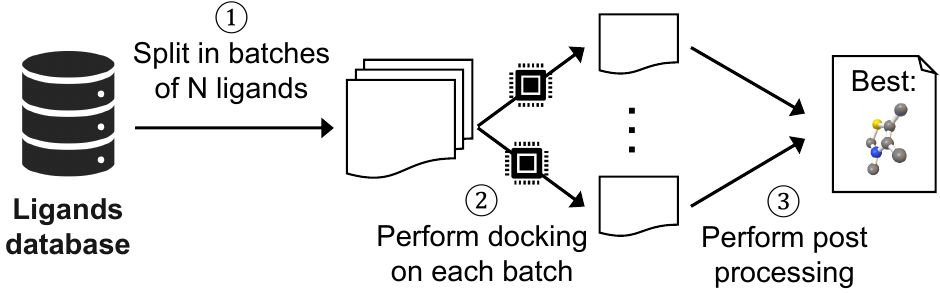}
    \caption{Virtual screening workflow. The ligand dataset is split in fixed-size batches~(\textcircled{\raisebox{-0.9pt}{1}}). Then, workers perform docking independently on each batch~(\textcircled{\raisebox{-0.9pt}{2}}). The results are then gathered for all batches, and post-processed to extract relevant domain-specific information~(\textcircled{\raisebox{-0.9pt}{3}}).}
    \label{fig:workflow_batch}
\end{figure}
A key challenge in virtual screening is to evaluate a very large number of drug candidates -- on the order of several millions of molecules. A simplified description of a virtual screening job is presented in Fig.~\ref{fig:docking_is_expensive}. The computational cost of this process is linear with the number $N$ of ligands to evaluate. A straightforward approach to perform this large-scale evaluation would be to perform docking as many times as there are ligands. However, this approach would be highly inefficient, since launching AutoDock-GPU comes at a cost, notably induced by the initialization of the CUDA runtime, and by the reading of the receptor file, which is constant across all runs. In addition, the cost of scheduling and starting tasks on any scheduler -- be it Kubernetes, Airflow, or Slurm -- is generally not negligible.

To solve this issues, a previous work~\cite{legrand_gpu-accelerated_2020} used a batching strategy for large-scale runs of AutoDock-GPU on the Summit supercomputer. This strategy first splits the ligand database into several fixed-size batches. For each batch, an index file contains a reference to the file describing the receptor molecule, and a list of all ligands files, which were obtained from splitting the ligand database. This batching strategy is supported in AutoDock-GPU, and we use it in for our virtual screening workflow. We illustrate this approach in Fig.~\ref{fig:workflow_batch}. As their workflow targets HPC environment, they used a custom scheduling mechanism to perform docking on the batches -- a fixed amount of workers are launched using a SLURM script, and each idle worker pulls a list of batches to process from a Redis database.
Different from the previous workflow~\cite{ornl_workflow} on HPC systems, we instead rely on the scheduling capabilities of Airflow to execute docking on all batches. For this purpose, we developed a DAG that fully describes our workflow requirements, including performing docking on multiple batches, where the processing of each batch is independent of other batches.

\subsection{Implementation}

\begin{figure}
    \centering
    \includegraphics[width=0.9\textwidth]{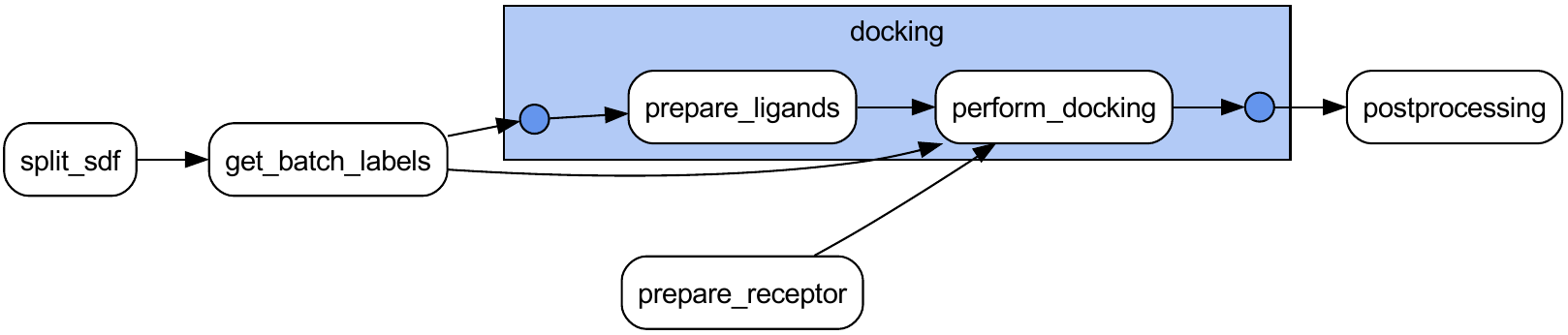}
    \caption{Our DAG for the AutoDock-GPU workflow on Apache Airflow. The blue rectangle indicates a task group, whose tasks are executed once for each batch.}
    \label{fig:the_dag}
\end{figure}

Our DAG description of the workflow is presented in Fig.~\ref{fig:the_dag}. Several tasks are defined to achieve our batched approach. First, the \verb|split_sdf| task creates several fixed-size batches from a single input file, which contains all ligand molecules to evaluate, this task returns the number of created batches. This number is then used by the \verb|get_batch_labels| tasks to generate a list of unique batch labels. We then use an Airflow feature, \textit{dynamic task mapping}, to instantiate a group of tasks for each batch in the batch list. This is represented in the blue rectangle on the DAG: for each batch, the task group \verb|docking|, which contains the \verb|prepare_ligands| and \verb|perform_docking| tasks, is instanced. Each task group instance takes a single batch label as parameter, and perform docking for this batch. For each batch, the ligands are first transformed by the \verb|prepare_ligand| task, which converts file formats, and transform ligand molecules. Then, the \verb|perform_docking| task uses AutoDock-GPU to perform the molecular docking job, this is the core computational task in our workflow. The \verb|postprocessing| task is executed when all batches have been processed, and performs gathering of results to provide domain scientist with relevant results, and visualization. In this DAG, parallelism is achieved by running concurrently several instances of the \verb|docking| task group.

{}

We execute the various tasks as Kubernetes pods. For this purpose, we created a single Docker container image that is used for all tasks. This image contains all tools required to run the tasks: AutoDock-GPU, various AutoDock preparation scripts, OpenBabel for molecule file format conversion, and autogrid4 for receptor pre-processing. The CUDA runtime is also included to enable running AutoDock-GPU on GPU hardware. Finally, we created a shell script for each task. Each script takes runtime parameters -- such as batch label and receptor file location -- as arguments, reads data from the file system, and writes results to both the file system and the standard output. We use our own file naming convention to ensure consistency between script execution. For example, \verb|db_batch35_ligand42.pdbqt| is the ligand N\textsuperscript{o}42 of the batch N\textsuperscript{o}35, from the ligand database \verb|db|.

In Airflow, tasks can be defined using \textit{operators}, which are task templates. In our DAG, we define most tasks with the \verb|KubernetesPodOperator|, which allows launching a pod in a Kubernetes cluster. To use this operator, we define the characteristics for the Kubernetes pod, using the standard pod specification format, as defined by the Kubernetes API. For all tasks, our pod specification comprises a reference to our custom Docker image, along with a task-specific shell command. In addition, we attach a persistent volume to each pod, at the \verb|/data| mount point in the containers, which we also use as the working directory for the container definition. This file system is used by our shell scripts to read and write data for the docking workload, and is shared between all pods. As we use the standard Kubernetes approach to attach storage to pods, the underlying storage technology and location are abstracted and can be easily modified to fit specific Cloud provider offers. To enable GPU-acceleration of the \verb|perform_docking| task, we define an additional pod specification for this task, which reuse the generic pod specification, with an added requirement for an NVIDIA GPU.

To enable communication between tasks, we use \textit{XComs} (short for ``cross-communication''), which is an Airflow-specific mechanism that allows tasks to communicate with each others. In particular, it can be used to pass parameters to tasks, along with collecting return values, when relevant. For instance, we used XComs in our DAG to collect the result of \verb|split_sdf|, and pass it to \verb|get_batch_labels|, along with passing batch labels to the task instances in the \verb|docking| task group. As a side note, XComs are not represented directly in the DAG, but are explicitly defined in the task definitions, as template strings. The values for XComs are unknown when the python code of the DAG is first parsed by Airflow, but they are populated as the DAG is being executed.

\section{Evaluation and Results}
In this section, we first study the performance impact of containerized environment on the GPU-accelerated docking. We then evaluate our workflow on a real-world ligands dataset, and study its resource utilization and concurrent task execution abilities.

\subsection{Evaluation setup and Datasets}
Our testbed is composed of a single server featuring a consumer-range CPU, along with a single low-end GPU. We deployed a Kubernetes cluster using the lightweight k3d\footnote{\url{https://k3d.io/}} distribution. We performed a full run of our workflow using this setup. We arbitrarily chose to perform docking on the \textit{Carboxypeptidase A} protein as receptor, with ligands from the SWEETLEAD~\cite{novick_sweetlead_2013} dataset, which contains approximately 10,000 chemical compounds and is widely used in drug discovery works. We used a batch size of 1,000 ligands per batch. This run is not relevant from the perspective of concurrent task execution, as only one GPU is available at a time, and thus only one docking task can be executed at a time. However, it allowed us to assess the correctness of our DAG and obtain some baseline measurements for future larger-scale executions. 

The total runtime for this job was $\sim$44~hours, that is, on average approximately 17 seconds to perform docking for one ligand. During this experiment, we observed a strong imbalance in processing time between batches, as some batches were processed in 30 minutes, while some others required 12 hours to finish. We suspect that this imbalance is caused by the distribution of molecule sizes within the original ligand database: as larger ligands may be grouped in specific regions of the file, batches containing those regions may be more computationally expensive to process.

\subsection{AutoDock-GPU in Containerized Environment}
\begin{figure}
    \centering
    \includegraphics[width=0.8\linewidth]{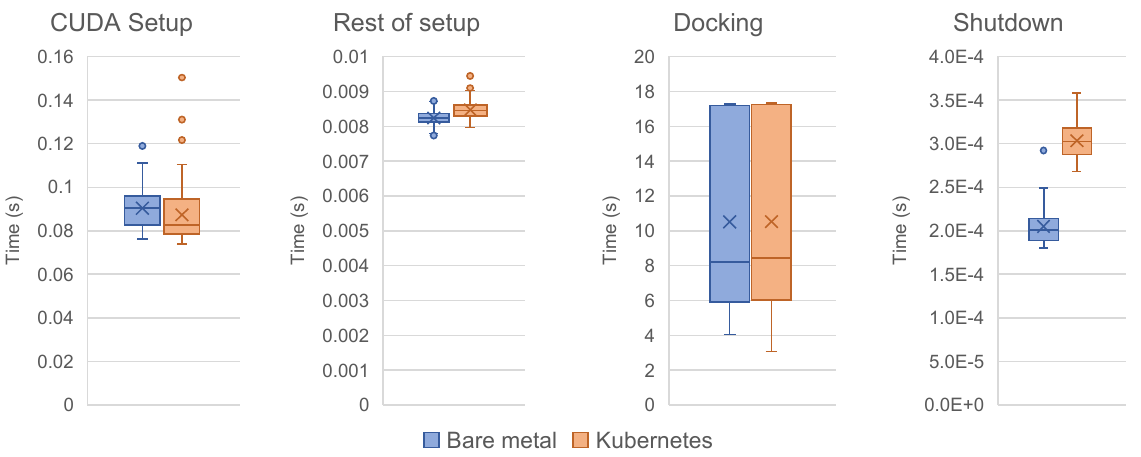}
    \caption{Performance comparison of AutoDock-GPU on Kubernetes/Airflow with execution on bare metal.}
    \label{fig:container_baremetal}
\end{figure}

Fig.~\ref{fig:container_baremetal} reports the distribution of the time measurements for 100 runs of AutoDock-GPU using a single protein-ligand complex, identified by the \texttt{7cpa} PDBID. Execution time for four phases is measured, including CUDA setup, rest of setup, docking, and shutdown -- those phases are reported by the AutoDock-GPU program, and refers to various phases of the program. We compare the execution time on the Kubernetes-Airflow setup with its equivalent bare metal execution. In order to ensure a fair comparison between both configurations, we use the same initialization seed for the pseudo-random number generator between the two methods. As the cloud environment often has other workloads co-running, we present the distribution of the execution time in a whiskers chart showing min, max, median, 25\%, and 75\% quantiles. The most significant difference lies in the shutdown phase, where the bare metal is much faster than the Airflow mode. However, in the dominant phase -- the docking process, both the bare metal and the Airflow execution have similar runtime with the bare metal exhibiting slightly lower time. The results show that for the main GPU-accelerated computation phase, deployment on Airflow/Kubernetes is feasible and performance comparable, likely because the GPU resource is not shared and thus not much influenced by other co-running workloads.

\subsection{Task scheduling and parallel execution}

\begin{figure}[bt]
    \centering
    \includegraphics[width=0.5\textwidth]{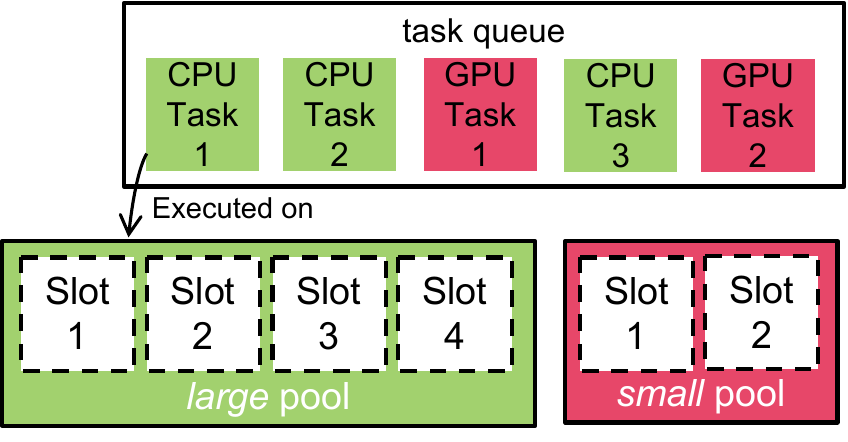}
    \caption{Tasks execution on Airflow pools. A task is associated to a specific pool, and a limited number of tasks associated with the same pool can run in parallel.}
    \label{fig:pools}
\end{figure}

To evaluate our workflow from the perspective of parallel task execution, we duplicated our real-world fully-functional DAG into a dummy DAG where the execution time of each task is controlled. In this DAG, no GPU resources are requested in the Kubernetes pod description, so that several docking tasks can be executed in parallel on our single-GPU setup. To ensure that this setup is still realistic, we enforced a limit on the number of concurrently running tasks, to simulate an environment where resources -- GPU and CPU -- are limited. To achieve this, we used the \textit{pool} feature of Airflow. Pools are used to limit the execution parallelism of a determined set of tasks. Fig.~\ref{fig:pools} shows a diagram to describe this concept. In our setup, we define two pools: a \textit{large} pool, which represents CPU resources, and a \textit{small} pool, which represents more expensive GPU resources. The \verb|perform_docking| task, which is the only task that uses GPU resources, is associated with the small pool, while all other tasks are associated with the large pool. Here, it is important to note that pools do not necessarily represent actual resources, but instead arbitrary limits.

\begin{figure}[bt]
    \centering
    \includegraphics[width=0.85\textwidth]{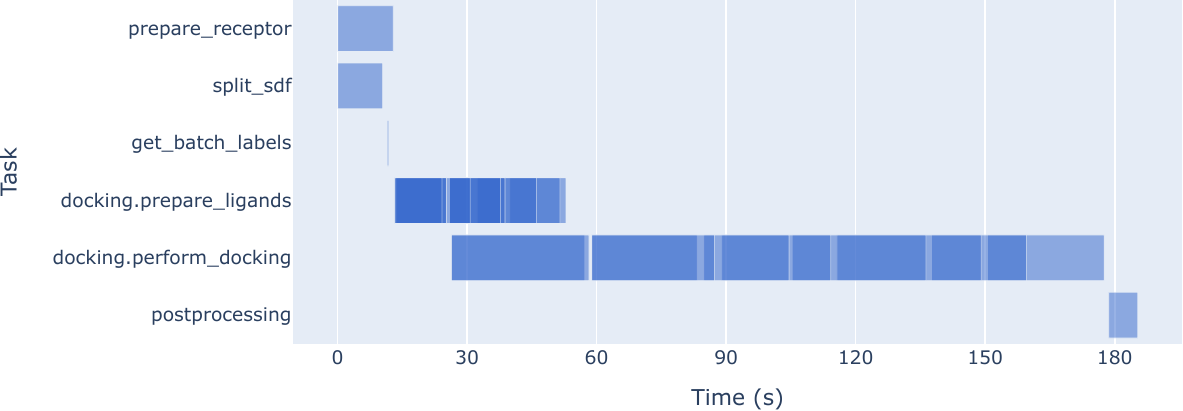} 
    \caption{Gantt chart of the execution of our dummy DAG with 10 batches in Apache Airflow. Task instances are plotted as translucent blue rectangles, darker blue indicate overlap of several rectangles, indicating parallelly executing task instances.}
    \label{fig:gantt_tasks}
\end{figure}

We run this dummy workflow using this two-pool configuration, with 4 slots in the large pool, and 2 slots in the small pools. The duration of each task instance is chosen randomly at runtime, and is on the order of several seconds, with the docking tasks set to take significantly longer than the other tasks. We chose to simulate the processing of 10 batches. Fig.~\ref{fig:gantt_tasks} presents a Gantt chart of this experiment, as found in Airflow's user interface, with improved visualization by using translucent colors. On this plot, each row represents a task, and each rectangle represents a task instance, the width of a rectangle represents the duration of the associated task instance. On this plot, we first observe that two tasks with no interdependence, such as \verb|prepare_receptor| and \verb|split_sdf|, are executed in parallel. The dependencies between tasks also naturally appears on this representation, as all instances of \verb|prepare_ligands| wait for the end of \verb|prepare_receptor| execution before starting. We also observe that some \verb|prepare_ligands| instances overlap with some \verb|perform_docking| instances. This happens naturally, as for a particular batch, the associated \verb|perform_docking| task instance only depends on the \verb|prepare_ligands| instance for this particular batch. These observations show that we achieved parallel task execution, with a simple DAG description of the workflow, and no custom scheduling logic.

\begin{figure}[bt]
    \centering
    \includegraphics[width=\textwidth]{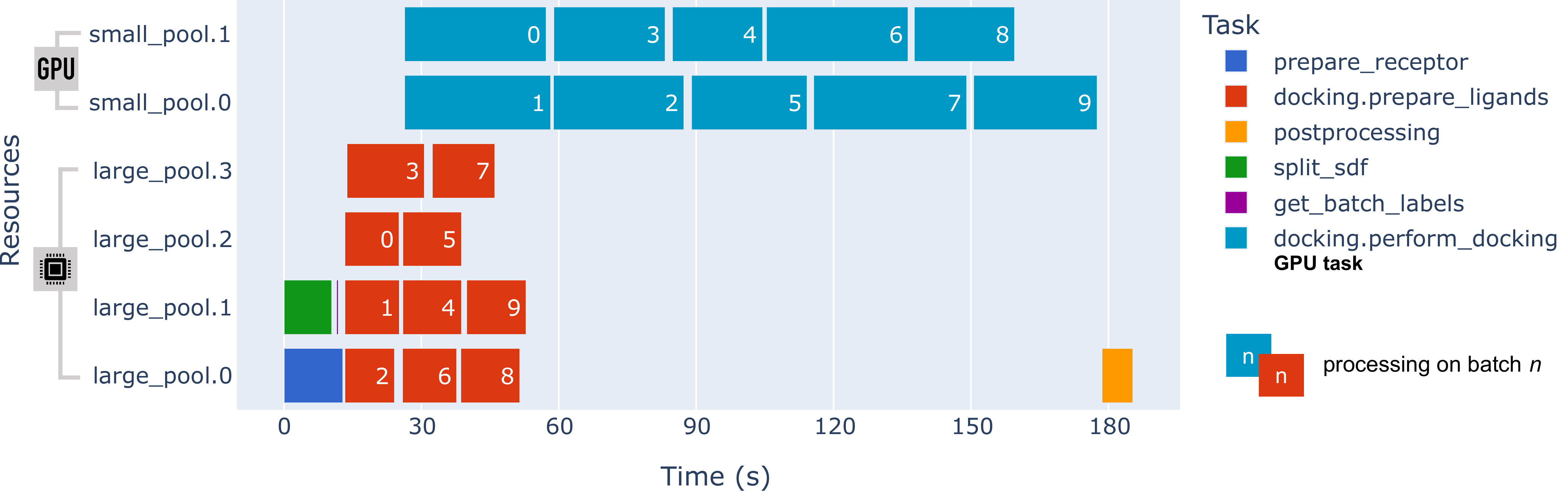}
    \caption{Resources-oriented Gantt chart of the execution of our dummy DAG with 10 batches, using two pools -- \textit{small} with 2 slots, and \textit{large} with 4 slots. Each line in this chart represents the activity for one slot. White numbers refer to the batch number associated with a task instance, when applicable.}
    \label{fig:gantt_resources}
\end{figure}

To further understand how tasks are mapped to resources, we propose in Fig.~\ref{fig:gantt_resources} a resources-oriented Gantt chart for the same experiment. On this chart, each line represents the utilization of a particular pool slot over time. It is worth noting here that this may not reflect actual resource utilization, as when a task start executing in Airflow, a pod creation request is submitted to Kubernetes, which then handles resources allocation. The execution of a task by Airflow is only conditioned by the availability of a pool slot; a task marked as ``running'' in Airflow may fail if Kubernetes is not able to meet the resource requirements for this task. 
On this chart, we visualize overlap in task execution, both between different tasks, such as \verb|split_sdf| and \verb|prepare_receptor|, and also between same-type tasks, but associated with different batches.
We also observe that GPU-enabled computing resources are only used to run GPU-accelerated tasks, which was a key motivation in this work, as those resources are quite expensive and should be efficiently utilized.
In addition, this Gantt chat highlights that the Airflow scheduler was able to provide with efficient scheduling of our tasks, given the requirements and interdependencies described in the DAG. As the Airflow scheduler is designed to accommodate several parallel DAG executions, other DAG-expressed workflows could easily be executed in this configuration. This would not cause any interference with our workflow execution, and would not require any modification to our DAG.


\section{Related Works}
\textbf{Scientific workflow platforms} Main frameworks are specifically designed for scientific workflows on HPC systems, including Pegasus~\cite{Deelman2005,Deelman2016}, Taverna~\cite{Wolstencroft2013}, FireWorks~\cite{Jain2015}, RADICAL-Pilot~\cite{Merzky2015}, Nextflow~\cite{DiTommaso2017}, signac~\cite{Adorf2018}, and CRCPs~\cite{Rosa2021}. Additionally, Pegasus and Nextflow can also be deployed in cloud environments; the former can be deployed on the cloud as a HTCondor instance while the latter has support to other platforms such as Kubernetes, Azure Cloud and Google Cloud. Computational Resource and Cost Prediction service~\cite{Rosa2021} allows users to control the financial costs of workflow execution on federated clouds.
\linebreak\linebreak \noindent
\textbf{HPC workloads in cloud environments} Several previous works have explored executing HPC workloads in cloud environments. Saha, et al.~\cite{saha2018evaluation} evaluated the Singularity containerization platform and Docker Swarm was used as an orchestrator, focusing on network mapping for MPI. 
Beltre, et al.~\cite{beltre2019enabling} run MPI workloads over TCP/IP and InfiniBand (RDMA) communication and measure the overheads between different container orchestrators.
Misale, et al.~\cite{misale2021towards} proposes a scheduler for Kubernetes (“KubeFlux”) based on the ideas from the Flux scheduler. Our work further expands the scope from a single HPC application but a workflow of multiple tasks on a cloud setting.

\section{Conclusions and Future Works}
The convergence of HPC and cloud computing is emerging to meet constantly evolving workloads. As a strong driver, complex workflows can benefit from efficient workflow management to ease the barrier to reaching wider user communities. Therefore, in this work, we investigated how a molecular docking workflow that was designed for HPC systems can be deployed on cloud-native infrastructure, represented by Kubernetes and Apache Airflow. We provide a design and implementation of a portable workflow description that supports parallel task execution on heterogeneous computing resources in Cloud environments. Our design batches a fixed number of ligands in one task to amortize overheads associated with Pod creation, termination, and I/O. We evaluated the workflow using a realistic dataset with ligands from the SWEETLEAD dataset. 
We find that predicting docking time based on ligand structures instead of simply the number of ligands may reduce load imbalance and improve scheduling efficiency. In our future works, we will also evaluate this workflow on a large-scale Kubernetes cluster along with elasticity support in the workflow.

\subsubsection{Acknowledgments} This research is supported by the European Commission under the Horizon project OpenCUBE (GA-101092984).

%
%
\bibliographystyle{splncs04}
\bibliography{main,wocc23}
\end{document}